\documentstyle[prl,aps,psfig]{revtex}

\newcommand{\eq}[2]{\begin{equation} #1 \label{eq:#2} \end{equation}}

\newcommand{\req}[1]{(\ref{eq:#1})}
\newcommand{\ave}[1]{\langle #1\rangle}
\newcommand{\rbt}[1]{\raisebox{1.5ex}[0pt]{#1}}
\newcommand{\qkl}[2]{{\large $\frac{{\rm #1}}{{\rm #2}}$}}

\begin{document}

\draft

\title{Landau Model for Commensurate-Commensurate Phase Transitions
       in Uniaxial Improper Ferroelectric Crystals}

\author{M. Latkovi\'{c} and A. Bjeli\v{s}}

\address{
 Department of Theoretical Physics, Faculty of Science \\
 University of Zagreb, Bijeni\v{c}ka 32, 10000 Zagreb, Croatia
 }

\author{V. Danani\'{c}}

\address{
 Department of Physics, Faculty of Chemical Engineering and Technology \\
 University of Zagreb, Maruli\'{c}ev trg 19, 10000 Zagreb, Croatia
 }

\maketitle

\begin{abstract}
We propose the Landau model for lock-in phase transitions in uniaxially
modulated improper ferroelectric incommensurate-commensurate systems
of class I. It includes Umklapp terms of third and fourth order and
secondary order parameter representing the local polarization. The
corresponding phase diagram has the structure of harmless staircase,
with the allowed wave numbers obeying the Farey tree algorithm. Among
the stable commensurate phases only those with periods equal to odd
number of lattice constants have finite macroscopic polarizations.
These results are in excellent agreement with experimental findings
in some A$_2$BX$_4$ compounds.

\end{abstract}


Many uniaxial systems undergo a sequences of incommensurate-commensurate
(IC-C) or commensurate-commensurate (C-C) phase
transitions~\cite{cumm,blinc}.
It is generally accepted that the main cause of such behavior comes
from the mode softening induced by competing interactions 
between neighboring sites in the crystal lattice~\cite{cowl,heine}.
The mode softening occurring at an arbitrary point in the Brillouin zone
characterizes the so called class I of IC-C systems,
in contrast to the examples of the class II for which the wave vectors of 
soft modes are at (or close to) the center or the border of the Brillouin 
zone~\cite{cowl}.

The IC-C and C-C phase transitions are mostly of the first order, and 
are accompanied with hysteresis and memory effects.
Also, depending on the wave number of modulation, some commensurate
phases have a uniform (e.g. ferroelectric) component. Recent experiments
show that the modulation of commensurate phases is usually domain like,
suggesting that multisoliton configurations that are usually present
in the incommensurate states, are to some extent frozen in lock-in phases 
as more or less dense soliton lattices~\cite{best,tsuda,hauke}.

Theoretical considerations of the class I of IC-C systems mostly start 
either from discrete microscopic models of competing interactions or from 
phenomenological expansions of the free energy functional. In the former 
cases the sequence of phase 
transitions is usually characterized by complete or incomplete devil's 
staircases for the wave number of ordering~\cite{aubr,janssen,neubert}. 
The crucial importance of discreteness in these models comes from the 
assumption that the couplings between neighbouring atoms or molecules
are very strong. However, as argued below, this assumption usually does 
not suit the microscopic properties of real materials.

The latter approach is based on the Landau theory of phase transitions, and
is generally appropriate for weakly coupled systems. The justification 
for continuous Landau models comes from many experimental indications, 
e.g. from the neutron scattering data~\cite{iizu,pate,iizu2,shig1}, showing 
a well-defined dispersion curves for collective modes with distinct soft-mode 
minima. The basic Landau model, commonly applied to various systems of the 
class I, includes one Umklapp term and the Lifshitz invariant. Within the mean 
field approximation it leads to the sine-Gordon problem~\cite{mcmi,bula,ishi} 
which has one isolated phase transition of the second order to the unique 
commensurate phase. However, in such systems one usually encounters more
commensurate phases, sometimes with the order of commensurability
much higher than three or four. 

The phases with high order commensurabilities can be explained only by
some extension of the sine-Gordon model. One widely explored way is to 
introduce an additional set of high order Umklapp 
terms~\cite{mashi,marion,parl3}, more precisely to take into account as 
many terms as there are stable commensurate states in the experimental
phase diagram. This approach is however not free from serious disadvantages.
At first, one introduces terms with powers of the order parameter
amplitude which are twice the order of commensurability, and as
such are beyond the standard (minimal) Landau scheme that includes
only those higher (usually fourth) order terms which necessarily
guarantee the boundness of the free energy density from below.
Further on and even more important, the microscopic analyses, which are 
usually avoided in such approaches, show that such terms are as a rule 
very small, and as such physically irrelevant in weakly coupled systems.

The model which is, in contrast to those of Refs.~\cite{mashi,marion,parl3},
formulated within the standard Landau scheme, and still explains the 
stabilization of high order commensurate phases, is proposed recently by 
two of us~\cite{bjel,lb}. In this model we take into considerations two
Umklapp terms of the lowest possible orders in the free energy
expansion, i.e. those of the third and fourth order, and so remain
strictly within the Landau theory~\cite{bc}. The similar starting point, 
but with Umklapp terms of orders higher than four, was proposed earlier in 
Ref.~\cite{folc}.
In Refs.~\cite{bjel,lb} we show that, due to the competition between these 
two Umklapp terms, the periods of ordering follow harmless 
staircase with the values obeying a Farey tree algorithm, while the 
corresponding commensurate configurations have the soliton lattice-like
forms. Also, the hysteresis and memory effects which are usually observed 
in particular materials, are within the present
model interpreted as an intrinsic property, i.e. as consequences of the
free energy barriers that appear due to the nonintegrability of the
Landau functional. These barriers prevent smooth transitions between 
neighbouring thermodynamically stable 
configurations~\cite{bari,kawa,dana,bjel,lb}.

The Landau model from Refs.~\cite{bjel,lb} contains only one order 
parameter, and as such does not possess all ingredients needed for the 
explanation of phase transitions in uniaxial improper ferroelectric 
materials. For such materials
it is necessary to include a secondary order parameter, that responsible 
for the net polarization of particular modulation structures. Although the 
necessity for such extension was put forward already by Iizumi et 
al.~\cite{iizu}, the subsequent studies of phase transitions in improper 
ferroelectrics were developed mostly within the simple sine-Gordon model 
with one Umklapp term~\cite{levan,shiba}. These analyses were 
concentrated on one isolated IC-C phase transition, allowing at most for its 
first order nature.

Our aim is to show that the model of Iizumi et al.~\cite{iizu}, extended with 
the additional Umklapp term which provides the mechanism for sequences of 
C-C phase transitions~\cite{bjel,lb}, enables the understanding of
the complex phase diagrams in A$_2$BX$_4$ compounds. In particular we explain 
several features of C-C phase transitions, like harmless staircase 
behaviour of the wave number of ordering, the first order nature of phase 
transitions, and the  polarization and modulation properties of commensurate 
phases.

We start from the assumption that the quadratic soft mode
contribution to the Landau expansion has minima at wave numbers 
$(+q_c,-q_c)$, where $q_4<q_c<q_3$, with $q_4=2\pi/4$
and $q_3=2\pi/3$ (the unit length is taken equal to the lattice
constant). The distances of $q_c$ from $q_3$ and $q_4$ are denoted
by $\delta_3$ and $\delta_4$ respectively,
with $\delta_3+\delta_4=\pi/6$, which gives the first independent
control parameter, $\delta_4$. The order parameter is complex,
$\rho e^{i\phi}$. Limiting the further analysis to the
temperature range well below the critical temperature for the
transition from the normal (disordered) to the incommensurate phase,
we also make the usual approximation of space independent amplitude
$\rho$, and keep only the phase-dependent part of the free energy density.

To allow for a finite polarization along some direction, we introduce 
a secondary order parameter $u$, and assume that it is coupled to 
the third-order Umklapp term, in agreement with the symmetry requirements 
for A$_2$BX$_4$ crystal lattice~\cite{iizu}. In order to stabilize the free 
energy with respect to changes in $u$, we also introduce the terms 
proportional to $u^2$ and $u'^2$. The complete free energy density reads
\eq{  f(\phi,u,x) = \frac{1}{2}\phi'^2 +
  \frac{1}{2}u'^2 + \frac{1}{2}\lambda u^2 + 
  Bu\cos{[3\phi+3(\frac{\pi}{6}-\delta_4)x]} +
  C\cos{(4\phi-4\delta_4 x)} , }{free}
where primes denote spatial derivation. The coefficients $\lambda$,
$B$ and $C$ in Eq.~\req{free} are rescaled in order to simplify equations. 
Thus $f(\phi,u,x)$ is the original free energy density divided by  
$\rho^2$. Note also that after this rescaling the coefficients $B$ 
and $C$ in Eq.~\req{free} are linear and quadratic 
in the order parameter amplitude $\rho$ respectively. Altogether, there are
three control parameters, namely $\delta_4$, $B$ and $C$, while the
parameter $\lambda$ just defines a scale for the polarization $u$.
Among the elastic terms in Eq.~\req{free}, the first one ($\phi'$-dependent)
favors the incommensurate sinusoidal ordering with the wave number $q_c$. 
The last two Umklapp terms cause the harmless staircase behaviour of the 
wave number of ordering~\cite{bjel,lb}.
The above Landau expansion follows, after spatial continuation, 
from microscopic models like e.g. those of Refs.~\cite{dbruce,cw}, that 
take into account local interactions presumably responsible for the mode 
softening (and for the coupling to the secondary order parameter), and start 
from the discrete presentation along the uniaxial direction.

The mean-field approximation for free energy~\req{free} leads to the
Euler-Lagrange (EL) equations:
\begin{eqnarray}
  & \phi'' & + 3Bu\sin{[3\phi+3(\frac{\pi}{6}-\delta_4) x]} + 
  4C\sin{(4\phi-4\delta_4 x)}=0 , \nonumber \\
  & u'' & - \lambda u - B\cos{[3\phi+3(\frac{\pi}{6}-\delta_4) x]}=0 . 
 \label{eq:el}
\end{eqnarray}
We are interested only in those solutions of EL equations which 
participate in the thermodynamic phase diagram of the model~\req{free}.
These solutions have, for given fixed values of the control parameters, 
the lowest values of the averaged free energy 
\eq{ \ave{F} = \frac{1}{L}\int dx f[\phi(x), u(x), x] .}{mfe} 
$L$ is the macroscopic length of the system.

Before calculating such solutions, let us make few remarks on EL 
equations~\req{el}. They can be considered as Lagrange equations for 
an equivalent classical mechanical problem. Note however that general 
("classical mechanical") solutions of these equations are not bounded, 
since $\lambda$ is positive. The only bounded solutions are 
periodic. From one side, obviously only such solutions may participate 
in the thermodynamic phase diagram. From the other side, the unboundedness
of all other solutions makes the analysis of the present model more
complicated than that of the model without the secondary order parameter 
$u$~\cite{bjel,lb}, for which the mechanical phase portrait is bounded, 
although chaotic~\cite{esca}.

It is clear from above remarks that periodic solutions are orbitally 
unstable. Therefore they cannot be calculated by a direct numerical 
integration of EL equations~\req{el}. Before embarking into another suitable
numerical method, it is useful to establish, by extending straightforwardly 
the previous treatment~\cite{lb}, necessary analytic conditions for these 
periodic solutions. We start by looking for allowed periods, after taking 
into account that periodic solutions have to obey the relations
\begin{eqnarray}
 \phi(x+P) & = & \phi(x)+\phi_P , \nonumber \\
 u(x+P) & = & u(x) ,
 \label{eq:phiper}
\end{eqnarray}
where $P$ is the period and $\phi_P$ is the phase increment per period.
From EL equations~\req{el} it follows that allowed periods and phase
increments are
\eq{ P=4k+3l, \hspace*{1cm} \phi_P=\delta_4 P-l\frac{\pi}{2} ,}{per}
where $k$ and $l$ are integers. The corresponding values of the total wave 
number (measured from the origin of Brillouin zone) are then
\eq{ 2\pi q \equiv q_c-\frac{\phi_P}{P} = 2\pi \frac{k+l}{4k+3l} .}{qkl}
     
\begin{figure}
\caption{Farey tree for wave numbers $q$ defined by
Eq.~\protect{\req{qkl}}.}
\centerline{\psfig{file=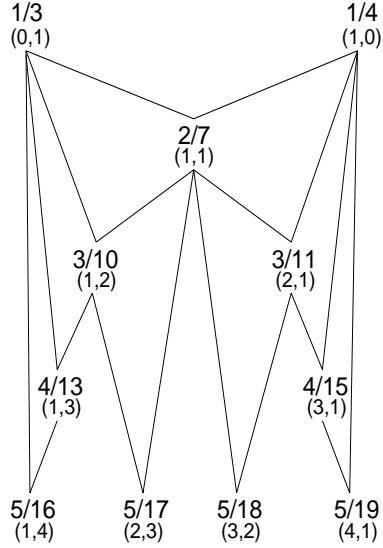,height=9cm,width=5cm,silent=}}
\label{farey}
\end{figure}

The values of $q$
given by Eq.~\req{qkl} form a Farey tree structure, shown in Fig.~\ref{farey} 
for wave numbers between $q=1/3$ ($k=0$, $l=1$ and $P=3$) and $q=1/4$
($k=1$, $l=0$ and $P=4$). The periodic solutions with $q=1/3$ and $q=1/4$ 
are the basic commensurate configurations, favoured by the 
Umklapp terms of third and fourth order respectively.
A higher order commensurate configurations are represented by
periodic solutions with wave numbers $q$ positioned
between $q=1/3$ and $q=1/4$ in the Farey tree. The values of integers 
$k$ and $l$ are positive for all these configurations. We emphasize that 
the Farey tree structure for $q$ is not the characteristic of usual 
continuous Landau models for IC-C transitions~\cite{mcmi,bula,ishi}.
Here it is imposed through the competition between two Umklapp terms in 
the first of EL equations~\req{el}.

For numerical purposes it appears convenient to eliminate the 
explicit $x-$dependence from one of Umklapp terms in the EL equations~\req{el}, 
by passing from the variable $\phi(x)$ to
\eq{ \psi(x)=\phi(x)+(\frac{\pi}{6}-\delta_4) x .}{psi}
The EL equations~\req{el} now read
\begin{eqnarray}
  \psi'' & + & 3Bu\sin{(3\psi)}+4C\sin{(4\psi-\frac{2\pi}{3}x)} = 0
  \nonumber \\
  u'' & - & \lambda u - B\cos{(3\psi)} = 0 .
 \label{eq:elpsi}
\end{eqnarray}
The corresponding free energy acquires a $\delta_4$-dependent term
in the form of Lifshitz invariant,
\eq{ F = \int dx\{ \frac{1}{2}[\psi'-(\frac{\pi}{6}-\delta_4)]^{2}+
        + \frac{1}{2}u'^2 + \frac{1}{2}\lambda u^2 + 
	Bu\cos{(3\psi)} + C\cos{(4\psi-\frac{2\pi}{3}x)} \} ,}{freepsi}
which considerably simplifies the calculation of the 
$\delta_4$-dependence of the averaged free energy~\req{mfe} for particular 
periodic solutions of the EL equations~\req{elpsi}. Note that these
equations do not contain the parameter $\delta_4$, so that it is sufficient 
to find solutions in variable $\psi(x)$ for some parameter values of $B$ and 
$C$.

Once the analytic conditions~\req{per} are established, the orbitally 
unstable periodic solutions of Eq.~\req{el} can be systematically calculated
by treating Eqs.~\req{el} as a boundary value problem, and using an 
appropriate numerical algorithm suitable for nonlinear differential 
equations, like the finite difference method.
The boundary conditions have to be specified on the left and right 
end points of the integration, i. e. for $x=x_0$ and $x=x_0+P$. It is 
convenient to put $x=x_0$ at one of the inflection points of 
$\psi(x)$~\cite{lb}. To complete the boundary conditions we still have to 
find out values of $\psi(x_0)$ and one of the values $u(x_0)$ or $u'(x_0)$,
as will be discussed in detail elsewhere~\cite{lbd2}. Tab.~\ref{table1}
contains choices of $x_0$, $\psi(x_0)$ and $u(x_0)$ or $u'(x_0)$ 
that specify completely the appropriate boundary conditions.
As previously~\cite{bjel,lb}, we find two independent
periodic solutions for given values of $k$ and $l$ (i.e. for a given 
wave number $q$). They differ by symmetry and have distinct
average free energies. These solutions are denoted as type A (antisymmetric
in $\psi$, symmetric in $u$) and type B (without particular symmetry).
These solutions are uniquely determined once the parameters 
$(k,l)$ [i.e. ($P,\phi_P$)] are chosen. Note that they do not depend
continuously on the parameter $\delta_4$. The additional minimization of the
free energy~\req{freepsi} provides only the $\delta_4$-dependence of ranges of 
stability for each pair $(k,l)$.

\begin{figure}
\caption{Solutions of EL equations~\protect{\req{elpsi}} $\psi(x)$
(phase of the primary order parameter) and $u(x)$ (local polarization):
$k=0$, $l=1$, $q=1/3$ (a), $k=1$, $l=0$, $q=1/4$ (b) and $k=1$, $l=1$,
$q=2/7$ (c). Other parameters are $B=0.1$, $C=0.1$, $\lambda=1$.}
\centerline{\psfig{file=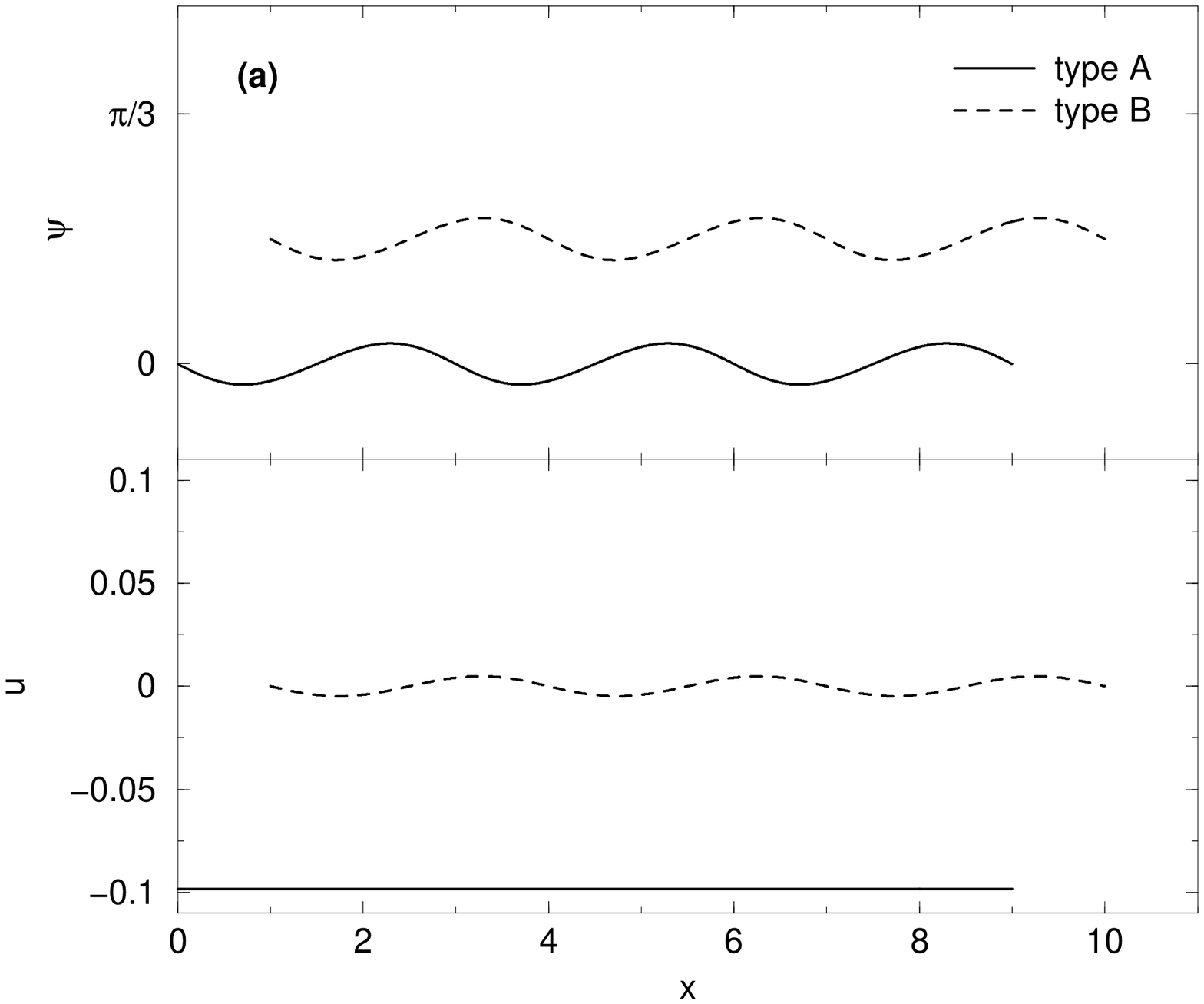,height=9cm,width=11cm,silent=}}
\centerline{\psfig{file=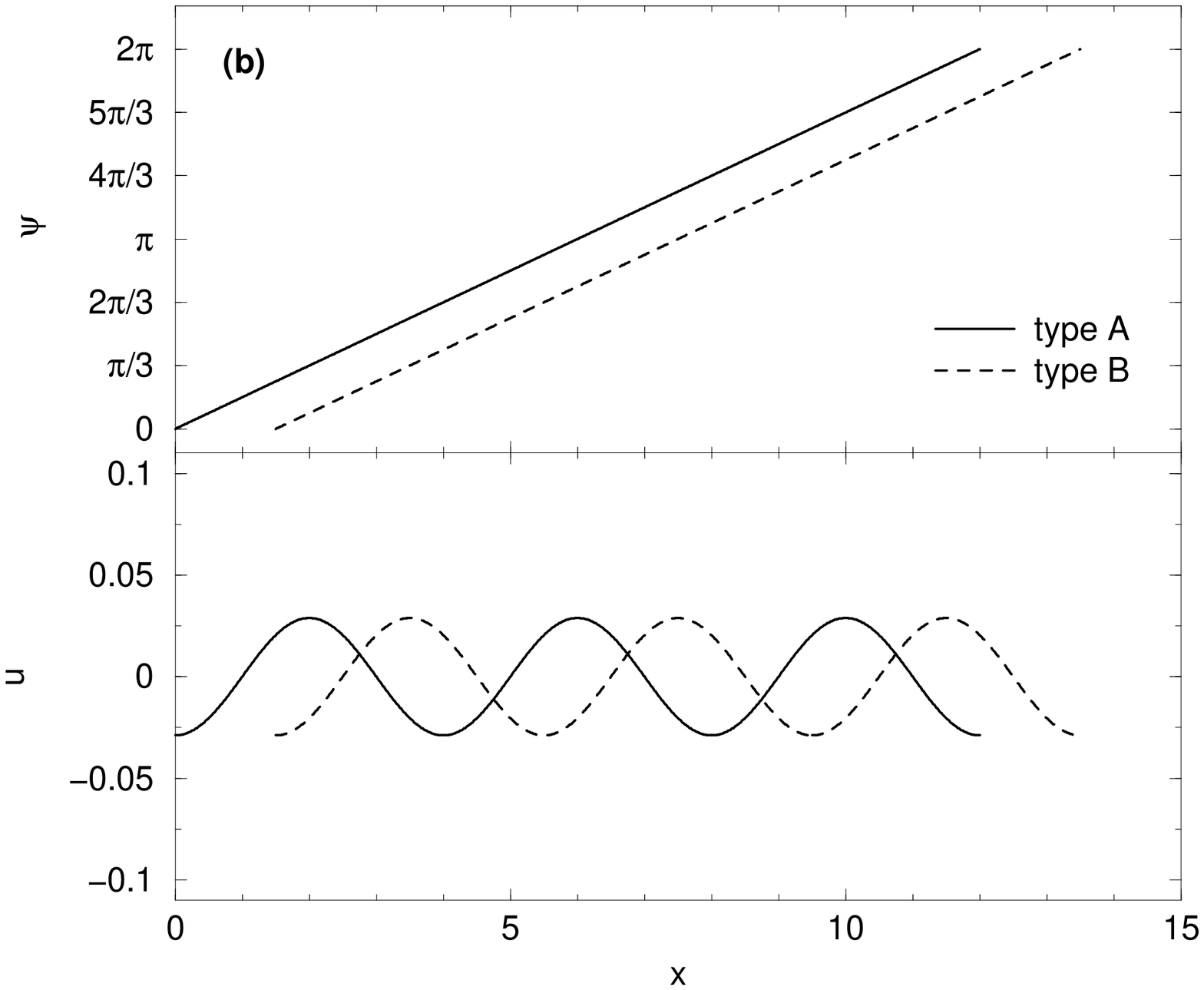,height=9cm,width=11cm,silent=}}
\centerline{\psfig{file=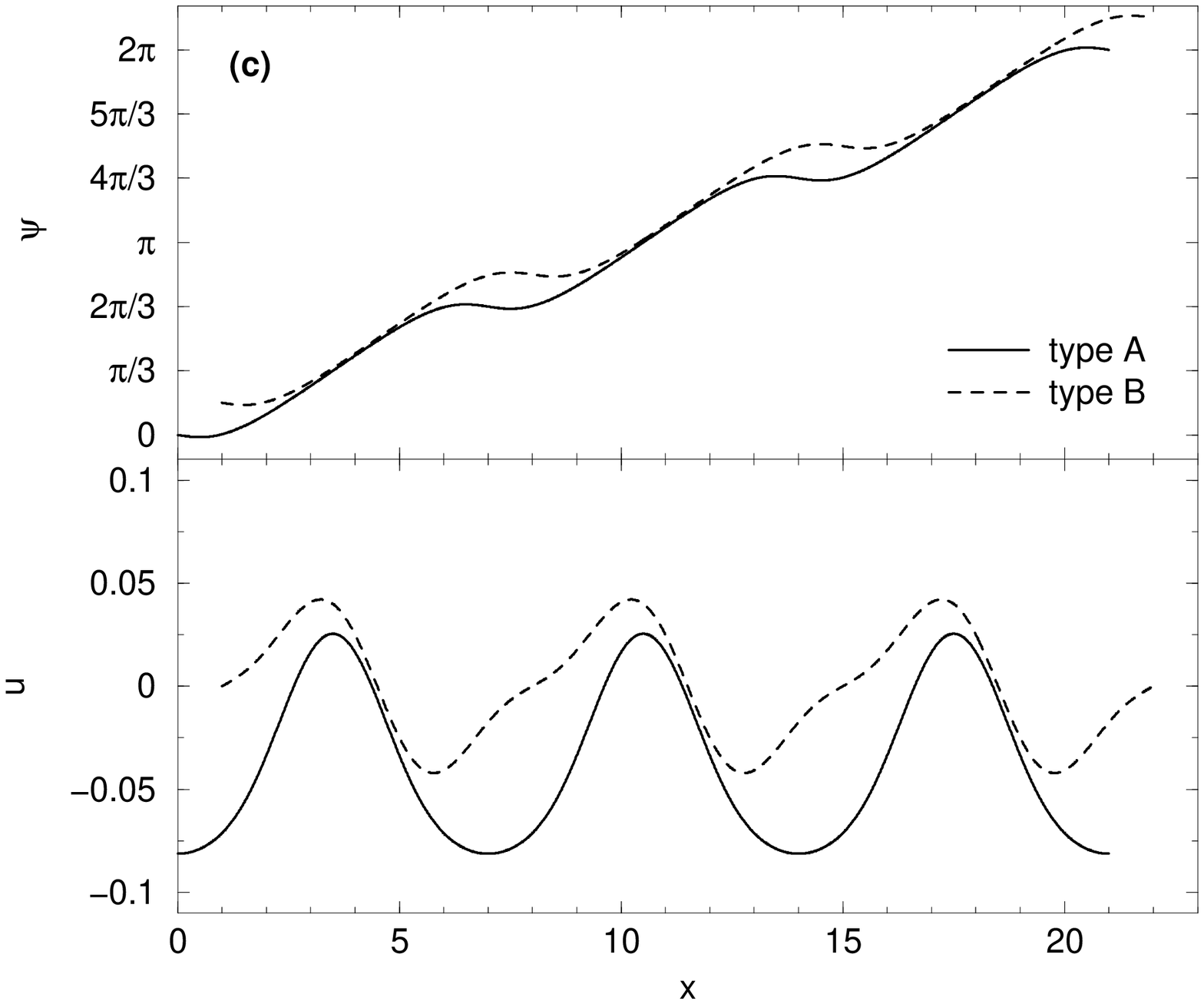,height=9cm,width=11cm,silent=}}
\label{persol}
\end{figure}

Three low order periodic solutions of Eq.~\req{elpsi}, with wave
numbers $q=1/3$, $1/4$ and $2/7$, are shown in Fig.~\ref{persol}.
For each of these values the phase of the primary order 
parameter, $\psi(x)$, shows qualitatively different beavior. A linear 
$x$-dependence of $\psi$ (i.e. a simple sinusoidal modulation of the 
primary order parameter) is realized for $q=1/4$. For $q=1/3$ $\psi$ has an
additional sinusoidal variation, while for  $q=2/7$ one encounters 
the periodic alternation of short domains with $q=1/4$ and $q=1/3$
modulations, or, in other words, the formation of a dense soliton lattice.
The tendency towards more and more dilute soliton lattices strengthens as
one goes down along the Farey treee from Fig.~\ref{farey}.
The secondary order parameter 
$u(x)$, i.e. the local polarization, either changes periodically in space 
with alternate positive and negative values forming an antiferroelectric 
lattice (type B solution from Fig.~\ref{persol}a, and type A and B 
solutions from Fig.~\ref{persol}b and \ref{persol}c), or has everywhere a 
ferroelectric space dependence (type A solution from Fig.~\ref{persol}a).

From the macroscopic point of view we are interested in spatially averaged
polarizations, given by
\eq{ \ave{u} = \frac{1}{P}\int_{x_0}^{x_0+P} u(x) dx }{aveu}
for periodic solutions. The values of $\ave{u}$ for solutions with 
various values of the wave number $q$ are listed in Tab.~\ref{table2}.
Note that the averaged polarization $\ave{u}$ vanishes for all solutions 
with even values, and all type B solutions with odd values, of the period $P$.

\begin{figure}
\caption{Phase diagram in the $(C,\delta_4)$ plane for $B=0.3$ and
$\lambda=1$. The numbers in the figure represents wave numbers of
ordering $q$ of commensurate phases. Phases with 
nonzero average polarizations $\ave{u}$ are marked by the
letter F.} 
\centerline{\psfig{file=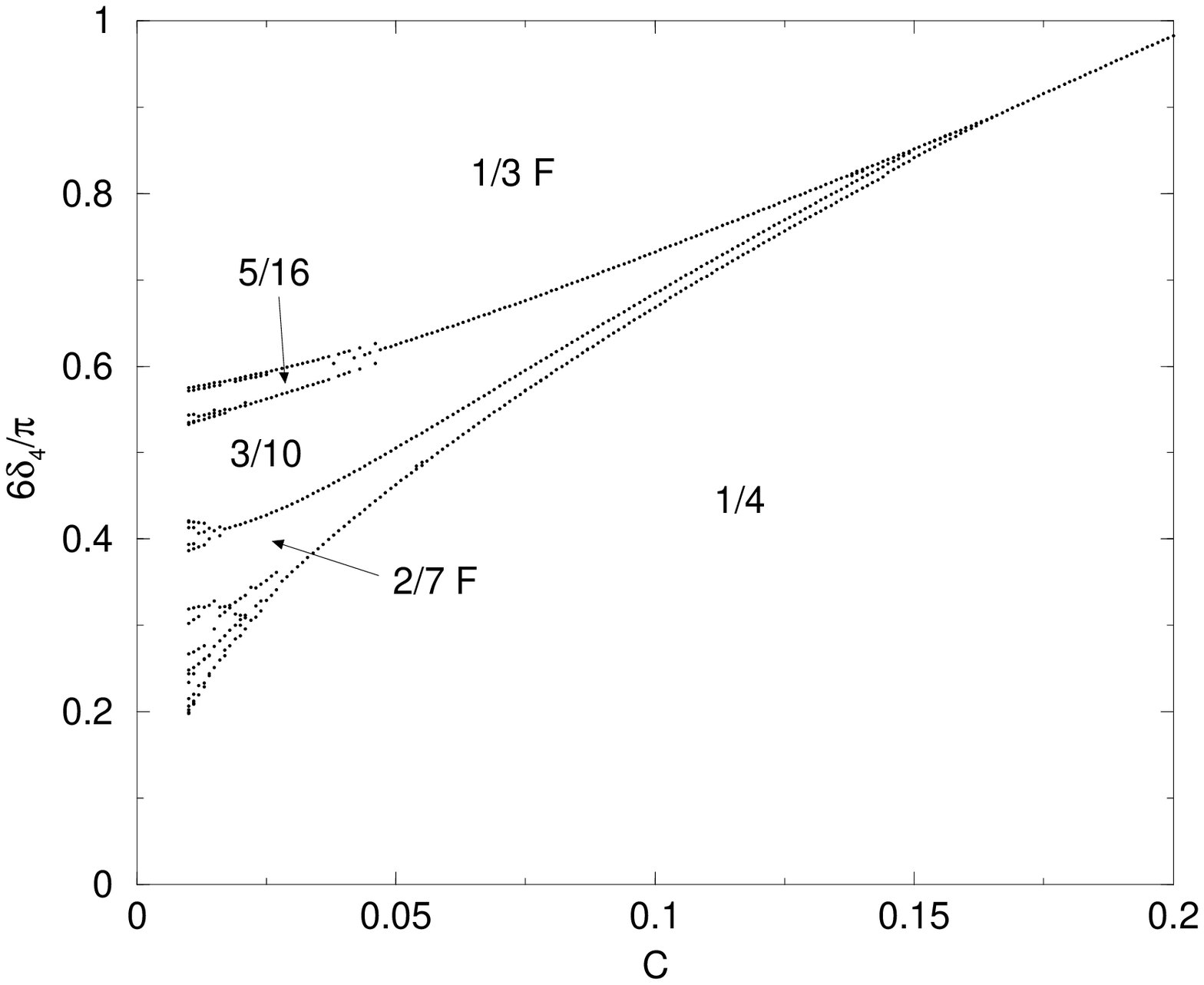,height=10cm,width=12cm,silent=}}
\label{diagram}
\end{figure}

The above boundary value method enables the calculation of thermodynamically
stable solutions for rather high values of the parameters $k$ and $l$
in Eq.~\req{per}. Here we limit the analysis of the phase diagram  
by keeping up to the tenth row in the Farey tree from Fig.~\ref{farey}. 
Also, in order to facilitate the further discussion, we fix
the value of the parameter B, and calculate the phase diagram in the reduced
parametric space $(C,\delta_4)$, as shown in Fig.~\ref{diagram}. Other choices 
of the cross sections in the parameter space~\cite{lb} lead to the 
qualitatively same conclusions.

As is seen in Fig.~\ref{diagram}, only the lowest order commensurate
phases, namely 1/3, 1/4, 2/7, and 3/10 are present for large values of
$C$. By decreasing $C$, i.e. by moving towards the sine-Gordon limit
($C\rightarrow 0$), more and more higher order commensurate phases start
to participate in the phase diagram. E.g., for $C \approx 0.05$ new
phases with the commensurabilities 3/11, 4/13, 5/16 are present.  

The phases which have nonzero average polarizations~\req{aveu} are denoted 
by the letter F in Fig.~\ref{diagram}. As numerical calculations
show, all phases with odd periods, e.g. those with wave numbers
1/3, 2/7 and 3/11, turn out to be ferroelectric. In other words, 
configurations with odd periods that 
have lowest averaged free energies, all belong to the type A solutions from 
Tab.~\ref{table2}. Note also that all lines in Fig.~\ref{diagram} represent 
phase transitions of the first order.

The phase diagram from Fig.~\ref{diagram} is in qualitative agreement with 
experimental phase diagrams for some members of the A$_2$BX$_4$ family. As
an example we take rubidium tetrabromozincate, Rb$_2$ZnBr$_4$, for which 
there is a variety of data on the temperature and pressure variation of the 
modulation wave number~\cite{iizu,pate,iizu2,parl1,parl2,shige}. Our phase 
diagram is in qualitative agreement with high pressure measurements collected 
in Fig.~9 of Ref.~\cite{parl1} and in Fig.~8 of Ref.~\cite{shige}.
In particular, the experimentally observed high order commensurate phases 
with $q$ equal to 2/7, 3/10, 3/11, 4/15, 5/17 and 7/24 are all present 
in Fig.~\ref{diagram}. Note also that our expression~\req{qkl} (and 
Fig.~\ref{farey}), which is obtained as an inherent property of the
competition between two Umklapp terms~\cite{comm}, represents the theoretic 
explanation of the phenomenological hint on the Farey tree structure of 
experimentally observed commensurabilities~\cite{parl1,parl2,shige}. Regarding 
the macroscopic polarization associated with various commensurate phases with 
odd periods, up to now the measurements are reported only for that with the
lowest commensurability (1/3) \cite{parl1,shige}. It shows a finite 
ferroelectric order, in agreement with our results. The experimental 
investigations of phases with higher commensurabilities, for which our 
results suggest presence or absence of a finite ferroelectric component, are 
highly desirable.

The next example from A$_2$BX$_4$ family of is ammonium
tetrachlorozincate, (NH$_4$)$_2$ZnCl$_4$. Its pressure-temperature phase
diagram~\cite{kity} is similar to that of Rb$_2$ZnBr$_4$. In particular
it contains a phase sequence with commensurabilities
1/3, 2/7, and 1/4, with some ambiguities about region of the stability of
phase 2/7 (see Ref.~\cite{sato} and references therein). The phase 1/3
is ferroelectric, the mixture of phases 2/7 and 1/4 shows a weak 
ferroelectricity, and some antiferroelectric order is observed in the 
nonpolar phase 1/4. This is in agreement with our results in Fig.~\ref{persol}. 
Indeed, the local polarization $u(x)$ for type A solution is constant for 
phase 1/3, it is nonpolar and has an antiferroelectric modulation for phase 
1/4, while for the phase 2/7 we expect a weak ferroelectricity since
the solution $u(x)$ of type A, although almost antiferromagnetically
modulated, still has a nonzero averaged value $\ave{u}$.

The last example, ammonium hydrogen selenate~\cite{deno}, NH$_4$HSeO$_4$ 
(together with its deuterated version ND$_4$DSeO$_4$), does not belong to the 
A$_2$BX$_4$ family. The sequence of phase transitions shown in Fig.~8 of 
Ref.~\cite{deno}, namely 1/3 (F), 3/10, 2/7 (F) and 1/4, can be also
reproduced by going along a particular path in Fig.~\ref{diagram}.
Let us also note that the present model is not directly applicable to 
betaine-calciumchloride-dihydrate (BCCD), a system with a well-known rich 
sequence of IC-C and C-C phase transitions~\cite{unruh,alme}. In this material, 
usually considered as a member of class II, one very probably encounters a 
competition of the first (ferroelectric, $q = 0$) and some higher order 
(probably $q = 1/4$) commensurabilities.

Let us finally shortly focus on the form of the commensurate modulations.
As already mentioned, multisoliton structures are sometimes frozen in 
commensurate phases well below the IC-C transitions~\cite{best,tsuda,hauke}.
Our numerical analysis shows that the stable solutions of higher 
orders (like that with $q=2/7$ shown in Fig.~\ref{persol}c) have the properties 
of dense, and not dilute, soliton lattices. It also suggests that particular 
solutions keep this form well below lock-in phase transitions, and that 
the polarization closely follows such behaviour of the dominant order
parameter by forming the modulated ferroelectric or antiferroelectric
patterns. 

In conclusion, by extending the basic model~\cite{bjel,lb} with the  
secondary order parameter representing the ferroelectric polarization, 
we obtain the phase diagram for the series of commensurate phases which is 
in qualitative and quantitative agreement with experimental phase diagrams
for some A$_2$BX$_4$ compounds~\cite{parl1,parl2,shige,kity,sato},
as well as for ammonium hydrogen selenate~\cite{deno}. The main
results of the present analysis are as follows. The wave number of ordering
is given by the harmless staircase obeying the Farey tree algorithm.
The commensurate phases are mainly characterized by dense soliton lattice
modulations. The phase transitions between successive lock-in phases
are of the first order. Finally, we establish the selection rule,
by which only those among the commensurate phases which have odd
periods are accompanied by finite macroscopic electric polarizations.

The work is supported by the Ministry of Science and Technology of 
the Republic of Croatia through the project no. 119201.


%
%

%
\begin{table}
\caption{Boundary conditions for periodic solutions of EL
equations~\protect{\req{el}}.}
\begin{tabular}{ccccccc}
 $k$      &  $l$     & Type& $x_0$ & $\psi_0$ & $u_0$ & $u'_0$ \\ \hline
          &          &  A  &  0    &    0     &       &  0     \\ \cline{3-7}
\rbt{odd} &\rbt{odd} &  B  &  1    & $\pi/6$  &   0   &        \\ \hline
          &          &  A  &  0    &    0     &       &  0     \\ \cline{3-7}
\rbt{even}&\rbt{odd} &  B  &  1    & $\pi/6$  &   0   &        \\ \hline
          &          &  A  &  0    &    0     &       &  0     \\ \cline{3-7}
\rbt{odd} &\rbt{even}&  B  & $3/2$ &    0     &       &  0     \\
\end{tabular}
\label{table1}
\end{table}

\bigskip

%
\begin{table}
\caption{ Polarization properties of commensurate phases with wave number
$q$ (Eq.~\protect{\req{qkl}}). Letter F indicates that particular solution
have nonzero average polarization (Eq.~\protect{\req{aveu}}).}
\begin{tabular}{ccccc}
$k$       & $l$      &      $q$            & Type & $\ave{u}$ \\ \hline
          &          &                     &  A   &     F     \\ \cline{4-5}
\rbt{odd} &\rbt{odd} &\rbt{\qkl{even}{odd}}&  B   &     0     \\ \hline
          &          &                     &  A   &     F     \\ \cline{4-5}
\rbt{even}&\rbt{odd} &\rbt{\qkl{odd}{odd}} &  B   &     0     \\ \hline
          &          &                     &  A   &     0     \\ \cline{4-5}
\rbt{odd} &\rbt{even}&\rbt{\qkl{odd}{even}}&  B   &     0     \\ 
\end{tabular}
\label{table2}
\end{table}

%
%

\newpage

\begin{references}
\bibitem{comm} Here the unit length is equal to the lattice constant
 in the uniaxial direction, in which negihbouring layers are connected by
 the glide plane symmetry element. Alternatively, particularly in theoretical
 works, the Brillouin zone is defined with respect to the interlayer distance
 (e.g. Refs.~\cite{cw,yh}).
\end{references}
\end{document}